\documentclass[preprint,showpacs,amsmath,amssymb]{revtex4}

\usepackage{graphicx}
\usepackage{dcolumn}
\usepackage{bm}

\begin{document}

\preprint{HUBP-08/02}

\title{An Analysis of Mutual Communication between Qubits by Capacitive Coupling}

\author{Takahiro Murakami}
\author{Masataka Iinuma}%
\author{Tohru Takahashi}
 \email[Corresponding Author: ]{tohrut@hiroshima-u.ac.jp}
\author{Yutaka Kadoya}
\author{Masamichi Yamanishi}
\affiliation{%
Graduate School of Advanced Sciences of Matter, Hiroshima University, 
Higashi-Hiroshima, 739-8530, Japan 
}%

\date{\today}

\begin{abstract}
A behavior of a two qubit system coupled by  the
electric capacitance has been studied quantum mechanically.
We found that
the interaction is essentially the same as the 
one for the dipole-dipole interaction; i.e., 
qubit-qubit coupling of the NMR quantum gate.
Therefore a quantum gate could be constructed 
by the same operation sequence for 
the NMR device if the 
coupling is small enough.
The result gives an information to the 
effort of development of the devices assuming 
capacitive coupling between qubits. 
\end{abstract}

\maketitle

 Toward the realization of quantum computers, 
various type of devices have
 been studied intensively in systems 
such as ion traps\cite{ion1,ion2}
, NMR\cite{NMR1}, linear optics\cite{lin1,lin2}, 
cavity QED with atoms\cite{cavqed1},
quantum dots in optical cavity
\cite{QD1,QD2}, and 
Josephson-Junction\cite{JJ1,NEC}, etc. 
In terms of the basic physics of the 
quantum gate, 
the quantum system
must satisfy
requirements that;
1) the transition between 
two levels in each qubit has to be
controllable independently for both phase and amplitude,  
2) qubits must have suitable mutual interaction
to construct a quantum gate.
It has to be emphasized that 
the first requirement means that the two-qubit system 
has to have a function to switch 
interaction between qubits during the 
quantum gate operation. 

Recently, an observation of Rabi oscillation 
in two-qubit system using the Josephson-Junction device
operating in the charge regime
has been reported\cite{NEC-new}.
It is an encouraging evidence of the existence
of the capacitive interaction between qubits.
However, the construction of the
universal quantum gate is yet to be demonstrated 
both theoretically and experimentally, i.e.,
a way of switching interaction between 
qubits is necessary to be demonstrated.
The switching  
could be realized by either an embedded mechanism in the device
or by a sophisticated operation with proper qubit-qubit
interaction.
The first one is simple and straightforward to understand, however, 
it may be difficult in the Josephson-Junction 
device in the charge regime or 
the Exciton-Photon device \cite{yamanishi-THZ,yamanishi-QIT}
since the device itself has to have a kind of 
mechanism to decouple two qubits.

For later case, a typical and a widely used way by 
NMR devices is to utilize the
dipole-dipole interaction between qubits which can be 
described by Hamiltonian as;  
\begin{eqnarray}
\begin{gathered}
  H_{\text{dipole}}  = \Omega _i \sigma _x^i  + \Omega _j \sigma _x^j  
   + \omega _i \sigma _z^i  + \omega _j \sigma _z^j  + \omega _{ij} \sigma _z^i \sigma _z^j  \\ 
\end{gathered} 
\label{eqn:ising}
\end{eqnarray}
where $\Omega _{i(j)}$ and  $\omega _{i(j)}$ are 
 the Rabi oscillation strength and the energy level of 
the quantum states
 in the i(j)-th qubit, 
while the last term describes dipole-dipole 
coupling between the i-th and j-th qubits 
with the strength of $\omega _{ij}$.
The Pauli matrix $\sigma _z^{i(j)}$ stands for the 
magnetic or the electric dipole operator
depending on the devices being considered.
  
The possibility of multi-qubits coupling 
using Josephson-Junction devices with the 
interaction (\ref{eqn:ising}) has been discussed
in \cite{JJ1} using a LC-oscillator mode coupling between
qubits, however, 
the Josephson-Junction 
device described in \cite{NEC-new} 
and  the Exciton-Photon device  
intend to use the capacitive coupling between 
two qubits rather than the dipole-dipole coupling.  
In these devices, 
the excited and the ground state in a qubit are
characterized by the difference of the electric charge
rather than the direction of the dipole moment  
so that the interaction  of the two qubits
is not described by
the same Hamiltonian as (\ref{eqn:ising}).

The operation of single qubit
by the Josephson-Junction  devices in 
the phase regime has been reported
\cite{JJ-phase}
and the quantum mechanical behavior of 
two-qubit coupling via capacitive coupling in the  
phase regime  has been 
studied by several authors\cite{Blais,Johnson,Strauch}.
These works showed the way to 
construct the universal quantum gate with 
the Josephson-Junction devices
in the phase regime.

In this letter, 
the quantum mechanical calculation of 
the behavior of the two-qubit system 
operating in the charge regime is reported.
We show that the system 
has the same nature with the dipole-dipole coupling, 
therefore, 
the same operation with the NMR devices
 are applicable to construct quantum gates 
in weak coupling regime

In order to see the behavior of the capacitive coupling
between two qubits,  
we analyzed the wave function of a two-qubit system
in four dimensional 
vector space;
$\psi  \equiv \varphi _1  \otimes \varphi _2 $,
where the basis of the space is defined explicitly as
\begin{eqnarray}
\left( {\begin{array}{*{20}c}
   {\left| 1 \right\rangle }  \\
   {\left| 0 \right\rangle }  \\

 \end{array} } \right)_1  \otimes \left( {\begin{array}{*{20}c}
   {\left| 1 \right\rangle }  \\
   {\left| 0 \right\rangle }  \\

 \end{array} } \right)_2  = \left( {\begin{array}{*{20}c}
   {\left| 1 \right\rangle \left| 1 \right\rangle }  \\
   {\left| 1 \right\rangle \left| 0 \right\rangle }  \\
   {\left| 0 \right\rangle \left| 1 \right\rangle }  \\
   {\left| 0 \right\rangle \left| 0 \right\rangle }  \\

 \end{array} } \right).
\label{eqn:base}
\end{eqnarray}
The time evolution  of each qubit
can be described by the 
Schrodinger's equation as  
$
i \hbar \frac{{d\varphi _i }}
{{dt}} = H_i \varphi _i $, where $H_i $
is the Hamiltonian of i-th qubit and its 
explicit form is, 
$H_i  = \left( {\begin{array}{*{20}c}
   {\Delta _i } & {a_i }  \\
   {a_i } & {-\Delta _i }  \\
 \end{array} } \right)
$
 with $\Delta _i$ and $a_i $
being the energy level and the Rabi oscillation strength of the qubit. 
It is assumed, as in all proposed devices, 
that Rabi oscillation in the qubit 
can be controlled by changing the energy level $\Delta _i$
via external parameters such as
voltages applied to the device.
Using these Hamiltonians,  the time evolution of the two-qubit system 
$\psi$ in four 
dimensional space is described as
$
i \hbar \frac{{d\psi }}
{{dt}} = \left( {H_1  \otimes I + I \otimes H_2  + H_{12} } \right)\psi 
$
where $H_{12} $
stands for the interaction between two qubits. 
As for the $H_{12} $,
we assume that 
the electric charge
appears only for the excited state 
so that an additional energy is put only when both 
qubits are in the excited state
($\left| 1 \right\rangle \left| 1 \right\rangle $ state in 
(\ref{eqn:base}))
\footnote{
In general, 
difference of the electric charge between the excited and 
the ground state is essential and the ground state is 
not necessary to be electrically neutral. 
However, it can be absorbed by the constant energy shift
in the diagonal element of the Hamiltonian without 
loosing accuracy of the discussion.}.
The Hamiltonian to describe the system is;
\begin{eqnarray}
H_{\text{cap}}  = \left( {\begin{array}{*{20}c}
   {\Delta _1  + \Delta _2 } & {a_2 } & {a_1 } & 0  \\
   {a_2 } & {\Delta _1  - \Delta _2 } & 0 & {a_1 }  \\
   {a_1 } & 0 & { - \Delta _1  + \Delta _2 } & {a_1 }  \\
   0 & {a_1 } & {a_1 } & { - \Delta _1  - \Delta _2 }  \\

 \end{array} } \right) + \left( {\begin{array}{*{20}c}
   {\Delta _{12} } & 0 & 0 & 0  \\
   0 & 0 & 0 & 0  \\
   0 & 0 & 0 & 0  \\
   0 & 0 & 0 & 0  \\

 \end{array} } \right)
\label{eqn:two}
\end{eqnarray}
where $\Delta_{12}$ in the second term is the coupling energy 
between two qubits. 
This assumption contrasts to the dipole-dipole interaction described in 
 Hamiltonian (\ref{eqn:ising}).
In fact, in  four dimensional vector space, 
 Hamiltonian (\ref{eqn:ising}) can be expressed as;
\begin{eqnarray}
H_{\text{dipole}}  = \left( {\begin{array}{*{20}c}
   {\omega _1  + \omega _2 } & {\Omega _2 } & {\Omega _1 } & 0  \\
   {\Omega _2 } & {\omega _1  - \omega _2 } & 0 & {\Omega _1 }  \\
   {\Omega _1 } & 0 & { - \omega _1  + \omega _2 } & {\Omega _2 }  \\
   0 & {\Omega _1 } & {\Omega _2 } & { - \omega _1  - \omega _2 }  \\

 \end{array} } \right) + \left( {\begin{array}{*{20}c}
   {\omega _{12} } & 0 & 0 & 0  \\
   0 & { - \omega _{12} } & 0 & 0  \\
   0 & 0 & { - \omega _{12} } & 0  \\
   0 & 0 & 0 & {\omega _{12} }  \\

 \end{array} } \right)
\label{eqn:nmr-four}
\end{eqnarray}
where the second term is a four dimensional expression of the 
dipole-dipole interaction.
We see the coupling energies are added symmetrically to the diagonal 
elements of the Hamiltonian and they also change their sign 
as the direction of a dipole flips. 
This feature plays an essential role to switch off the 
interaction between qubits effectively by the refocusing operation.

In order further to see characteristics of the capacitive coupling, 
it is useful to re-write the Hamiltonian (\ref{eqn:two}) in a two component form as;
\begin{eqnarray}
\begin{gathered}
  H'_{\text{cap}}  =   \frac{{\Delta _{12} }}
{4}I + a_1 \sigma _x^1  + a_2 \sigma _x^2 
   + \left( {\Delta _1  + \frac{{\Delta _{12} }}
{4}} \right)\sigma _z^1  + \left( {\Delta _2  + \frac{{\Delta _{12} }}
{4}} \right)\sigma _z^2  + \frac{{\Delta _{12} }}
{4}\sigma _z^1 \sigma _z^2  \\ 
\end{gathered} 
\label{eqn:four}
\end{eqnarray}

A comparison of the Hamiltonian (\ref{eqn:ising}) and (\ref{eqn:four})
clearly shows similarities and differences of the  two couplings schemes. 
Both of the two have the same type of the dipole coupling term as seen 
in the last term of Hamiltonians. 
On the other hand, in Hamiltonian (\ref{eqn:four}),
the coupling energy $\Delta_{12}$ are added to  
the energy level of  each state 
as shown in the fourth and the fifth term.
As a result, 
the energy level, $E_{1(2)}$, of the quantum state in the 1(2)-th qubit
can be expressed as;
\begin{eqnarray}
E_{1(2)}  = \Delta _{1(2)}  + \Delta _{12} /4 \pm \Delta _{12} /4
\label{eqn:level}
\end{eqnarray}
where the first two terms represent modified but fixed energy 
level of the quantum state in the qubit
while the last term is the contribution from the 
dipole type coupling in Hamiltonian (\ref{eqn:four}).
The sign of the last term depends on the relative states 
of the two qubits.
This fact shows that 
interaction of the capacitive coupling can be  
described essentially by 
the same form with the dipole-dipole couplings.
Therefore, 
it may be possible to perform 
the same quantum gate operations 
which have been applied on 
devices of the dipole-dipole coupling
such as NMR devices.

To realize the operation discussed above, the most important
condition on the parameters is the strength of the 
qubit-qubit coupling, $\Delta _{12}$.
Since the energy level of the quantum state in the qubit,$E_{1(2)}$, 
depends on the state of the neighboring qubit as expressed by the 
$\pm$ sign in the equation (\ref{eqn:level}), 
the quantum state of a qubit affects
the condition of Rabi oscillation of neighboring qubit.
This fact does not allow the independent control of 
each qubit. 
However, if the $\Delta _{12}$ is 
smaller enough than the $a_{1(2)}$,
the condition of Rabi oscillation 
can be  virtually independent of neighboring qubit 
as typical width of Rabi resonance 
is its  strength $a_{1(2)}$.

It has to be reminded that the situation described above is 
the same for the interaction with the Hamiltonian (\ref{eqn:ising});
i.e., for NMR devices.
The quantum operations demonstrated using the NMR devices
always have been performed in the weak coupling regime.
Since the spin-spin coupling of the NMR devices is so small that 
the  condition of weak coupling has been  satisfied 
without any special treatment.

It is now known that 
the capacitive coupling interaction also has
the dipole-dipole feature as shown in the Hamiltonian (\ref{eqn:four}), 
so that devices with capacitive  coupling 
could be operative as a universal quantum gate
utilizing the same operation sequence on the NMR devices.

\begin{figure}
\includegraphics[scale=0.8]{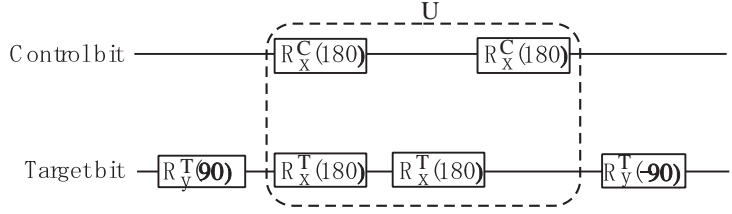}
\caption{\label{fig:opr} 
The controlled-NOT operation for a NMR device described 
in Ref.~\cite{NMR-CNOT}.
$R_i^{C(T)}(\theta)$ stand for $\theta$ rotation around the $i$ axis on 
the Control(C) or Target(T) bit. 
The first and the last $R^T_y(90)$ stand for Rabi oscillation of 90 degree 
while a series of operation denoted as U is a phase operation on the qubits, 
showing that the operation includes all components necessary to construct
the general quantum operations. }
\end{figure}

\begin{figure}
\includegraphics[scale=0.8]{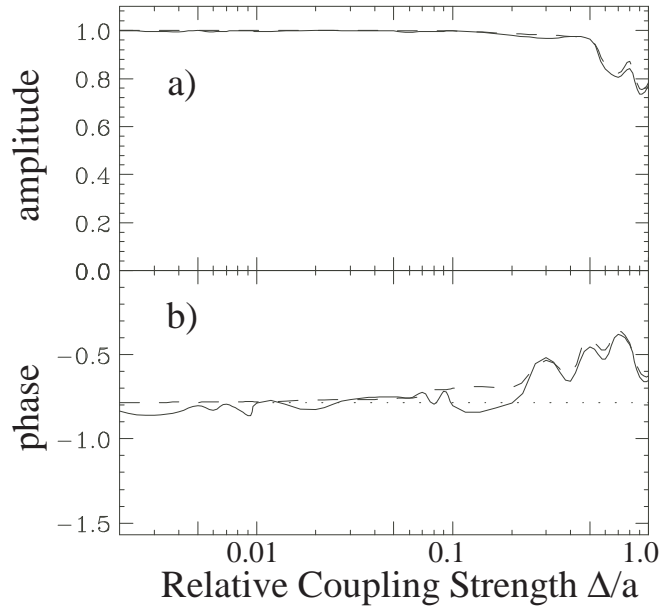}%
\caption{\label{fig:d12} 
$\Delta_{12}$ dependence of the amplitude (a) and 
phase(b) of
$\left| 1 \right\rangle \left| 0 \right\rangle $ state.
The expected value for the amplitude is 1.0 and for the phase is 
$-1/4\pi$(indicated by dots) respectively.
The solid line is for the case that 
Rabi oscillation strength, $a_{1(2)}$, is ON throughout 
the operation and the dashed line is for the case that 
$a_{1(2)}$ is ON only when the Rabi Oscillation stage.} 
\end{figure}

In order to see feasibility of 
the quantum computation with weak capacitive coupling, 
we performed a numerical calculation of a quantum gate operation using 
the Hamiltonian~(\ref{eqn:two}).
As an example of the two-qubit operation, 
we tried a controlled-NOT operation by the 
procedure commonly used in NMR devices\cite{NMR-CNOT}
which is 
schematically expressed in Fig.~\ref{fig:opr}.
It has to be noted that 
even though it is a single controlled-NOT operation,
it consists of all necessary operations 
for general quantum gate operations.
The controlled-NOT operation described in Fig.~\ref{fig:opr}
is expressed in term of a unitary transformation
as;
\begin{eqnarray}
\psi _i  \Rightarrow \psi _f = \sqrt { - i} \left( {\begin{array}{*{20}c}
   0 & 1 & 0 & 0  \\
   1 & 0 & 0 & 0  \\
   0 & 0 & 1 & 0  \\
   0 & 0 & 0 & 1  \\

 \end{array} } \right)\psi _i 
\label{eqn:cnot}
\end{eqnarray}
after subtracting overall phase factor. 
In the calculation, the initial state was chosen as
$\psi _i  = (1,0,0,0)$
on the basis shown in Eq.~(\ref{eqn:base})
and the expected final state is
$\psi _f  = (0,e^{-1/4\pi},0,0)$.
As the result of the calculation, 
we plotted, in Fig.~\ref{fig:d12},
the amplitude and the phase of the
$
\left| 1 \right\rangle \left| 0 \right\rangle 
$
state as a function of $\Delta_{12}$ normalized to 
$a_{1(2)}$,  
where we expect 1.0 and $-1/4\pi$ for the amplitude and 
the phase respectively.
In the  calculation, we also have to consider 
 treatment of 
the Rabi oscillation strength, $a_{1(2)}$, in a qubit.
Depending on the devices, 
we can assume that parameter $a_{1(2)}$
exists throughout the operation or appears only 
when the device is on Rabi oscillation.
Since it depends on devices considered, we
calculated both cases.

It can be concluded that the result are
reasonably close to the expected values and are
stable up to 
$\Delta _{12}  \approx 0.1a_{1(2)} $.
As for the treatment of $a_{1(2)}$, 
some deviation from the ideal value is 
seen if the $a_{1(2)}$ is ON throughout the calculations,
particularly for the phase of the sate.
It is preferable to control $a_{1(2)}$ as is realized
in NMR devices, however, 
the deviation appears to be
acceptable level.  
It is also worthwhile mentioning that 
 $\Delta _{12}  \approx 0.1a_{1(2)} $ is  much lager than 
those of typical NMR devices. 
It indicates that 
the devices with capacitive coupling potentially have 
advantage of  larger signal than those of
 NMR devices.

 In summary we analyzed the interaction between two charge qubits
via capacitive coupling.
We conclude that the coupling is essentially  
the same as
the dipole-dipole type interaction
and  the device can be a operative as a 
quantum gate if the coupling is
smaller enough than the
Rabi oscillation strength.
This fact gives an important information  for the
development of those devices such as Josephson-Junction 
or Exciton-Photon devices.

The authors would like to thank 
 late Prof. Osamu Miyamura of Hiroshima 
University who 
started this work. 
One of the author(T.M) specially 
 thank him for 
his guidance and encouragement.
This work is supported in part by 
the Grants-in-Aid for Scientific Research from Japan  Society for the Promotion of Science and 
Iwata Real-Time Recognition Project of
Hiroshima
Prefectural Institute of Industrial Science and Technology.

\end{document}